\documentclass[a4paper,11pt]{article}
\usepackage[utf8]{inputenc} 
\usepackage[T1]{fontenc} 
\usepackage{geometry}
\usepackage{setspace}
\usepackage[numbers, sort&compress]{natbib}
\setcitestyle{et al., maxauthors=3} 
\usepackage{authblk}
\usepackage[hidelinks]{hyperref} 
\usepackage{url} 
\usepackage{graphicx}
\usepackage[shortlabels]{enumitem}
\usepackage{caption}
\usepackage[scaled=1]{helvet}

\usepackage{pdflscape}
\usepackage{longtable} 
\usepackage{booktabs} 
\usepackage{lscape} 
\usepackage{array} 
\usepackage{multirow}
\usepackage{colortbl}
\usepackage{xcolor}
\usepackage{amsfonts} 
\usepackage{dsfont} 
\usepackage{microtype} 
\usepackage{lipsum} 
\usepackage{doi}
\usepackage{framed}
\usepackage{longtable}
\usepackage{ragged2e}
\usepackage{amsmath,amssymb}
\usepackage{calc}
\usepackage{mathtools}
\usepackage{fancyhdr}
\usepackage{siunitx}
\usepackage{xr}
\graphicspath{{/home/marc/Documents/RPROJ_2/CAD_stroke/reports/TEX/ARXIV/PICS/}}
\author[1,*]{Marc Delord}

\affil[1]{School of Life Course \& Population Sciences, Department of Population Health Sciences, King's College London, London, United Kingdom}
\affil[*]{Correspondence: marc.delord@kcl.ac.uk\\ Tel: +44 20 7848 8710}
\title{{Selection and Collider Restriction Bias Due to Predictor Availability in Prognostic Models}}
\date{}

\begin{document}
\maketitle

\setstretch{1.3}


\setstretch{1.3}

In medicine, prognostic scores or clinical prediction models are statistical models intended to estimate an individual’s probability of experiencing a specific health outcome over a defined period, based on their clinical and non-clinical characteristics \cite{altman2000we, Moons2009a, steyerberg2009applications, steyerberg2013prognosis}. Outcomes can refer to clinical events such as disease onset, complications, referral to secondary care, organ failure, or death. Classical examples of prognostic scores include the Framingham cardiovascular risk score, originally developed to estimate long-term coronary heart disease risk, and more recent tools such as the QRISK3 score, widely used in UK primary care to predict 10-year cardiovascular risk.



The inflation in the number of published prediction models over recent decades has generated an extensive methodological literature on common shortcomings in prognostic model development and guidelines for their design, validation, and reporting \cite{wyatt1995commentary, Moons2009a, royston2009prognosis, Altman2009, Moons2009b, hemingway2013prognosis, collins2015transparent, steyerberg2013prognosis, efthimiou2024developing}. In broad terms, the development of prediction models should include the development of the prediction model itself \cite{royston2009prognosis}, its external validation and calibration \cite{Altman2009, altman2000we}, and, ideally, the conduct of a prospective clinical and economic impact study \cite{Moons2009b}.


This well-defined framework, intended to ensure the methodological soundness of prediction model assessment implicitly assumes that required predictors are routinely available at the point of care \cite{Moons2009a,royston2009prognosis}, an assumption that, despite its importance, has received little explicit attention in the methodological literature on prediction models \cite{efthimiou2024developing}. More generally, when a prognostic score is developed or validated using retrospective data, inclusion is restricted--by construction--to patients with recorded predictors. This restriction to patients with recorded predictors is not inherently problematic, and if measurement occurs independently of determinants of outcome risk, the analysed sample may still yield unbiased estimates. However, when predictor measurement depends on underlying disease severity or related care processes, restriction selects individuals based on a variable influenced by determinants of the outcome. This process is analogous to protopathic bias, a form of revers causality in which early manifestations of disease prompt intervention before formal diagnosis \cite{feinstein1985clinical}.

\begin{figure}[!ht]
\centering
\includegraphics[scale = .5, trim = 0cm  0cm 0cm 0cm, clip]{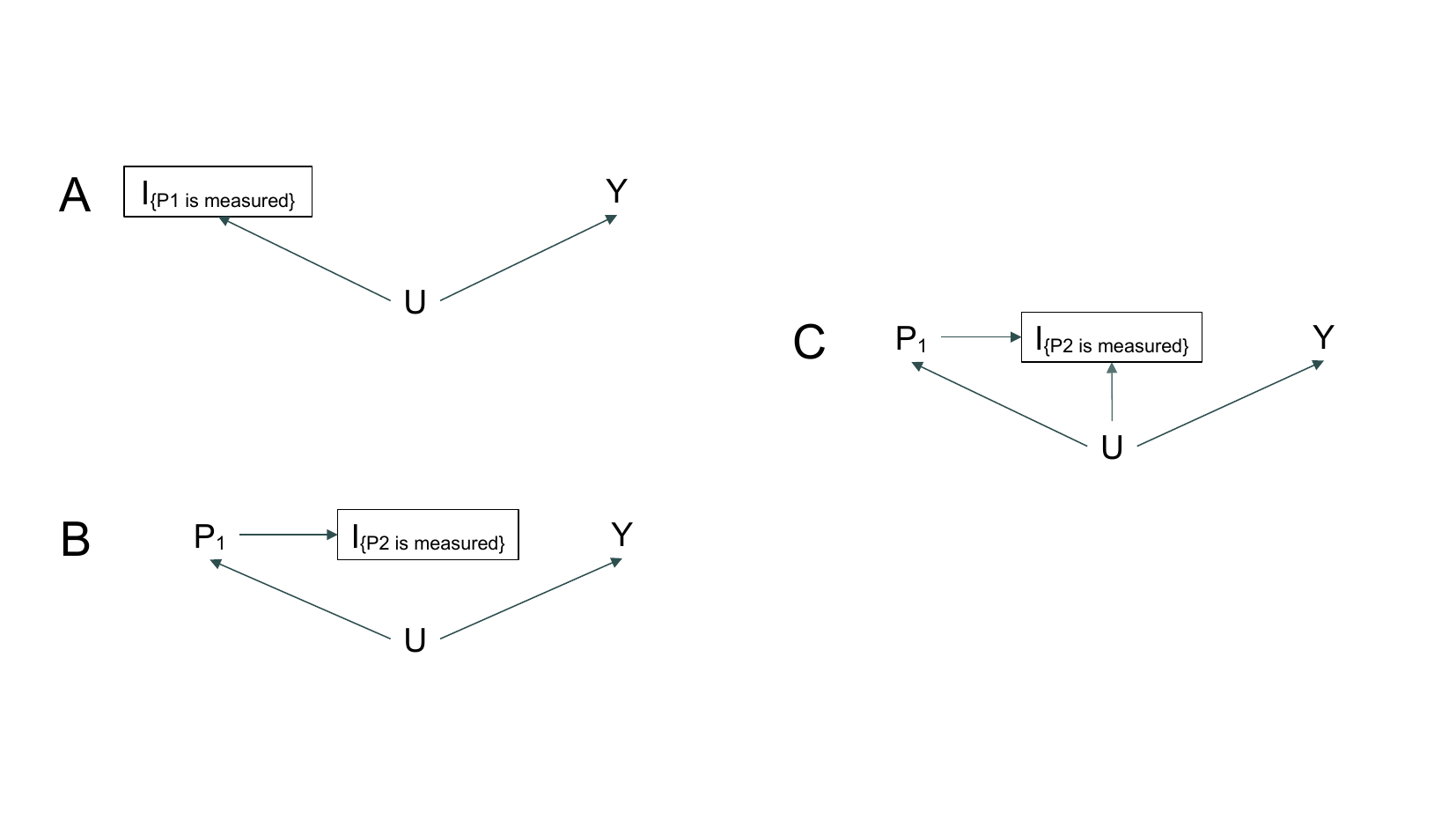}
\caption[]{Selection and collider restriction bias arising from conditioning on predictor availability in prognostic models. Directed acyclic graphs are used to illustrate causal dependencies between observed and unobserved variables. $U$ represents unobserved disease severity; $P_1$ and $P_2$ represent predictors; and $Y$ represents the outcome. Framed nodes indicate measurement-based restriction.
Panel A depicts simple selection bias, where underlying disease severity triggers measurement of the predictor. Panel B depicts simple selection bias in which disease severity is captured through a measured proxy. Panel C illustrates collider restriction bias, where both underlying disease severity and its proxy influence measurement of $P_2$, making its availability a collider.
}
\label{fig:1}
\end{figure}

This situation, illustrated in panel A of Figure \ref{fig:1}, corresponds to classical selection bias: underlying disease severity (U) influences both the outcome and the likelihood of predictor measurement so that restricting analysis to individuals with recorded predictors effectively selects on severity.\\

A concrete example illustrating how the assumption of unbiased predictor availability may not hold in practice is provided by the Kidney Failure Risk Equation (KFRE) \cite{tangri2011predictive}. The KFRE is a prognostic model developed to predict progression to kidney failure in patients with chronic kidney disease stages 3–5 (CKD 3–5), defined by persistent reduction in kidney function or markers of kidney damage. It estimates an individual’s risk of kidney failure at 2 or 5 years and is intended to inform risk stratification and referral decisions in patients with chronic kidney disease. The commonly used four-variable version relies on age, sex, estimated glomerular filtration rate (eGFR), and urine albumin-to-creatinine ratio (uACR), a measure of albuminuria. Although the KFRE has been the subject of numerous validation studies reporting strong predictive performance \cite{peeters2013validation, tangri2016multinational, Major2019, ramspek2020towards}, its uptake in routine clinical practice remains limited \cite{ramspek2020towards}. This limited use may be partly explained by constraints in routine data availability \cite{Fine2013}, with eGFR or uACR not being systematically recorded in community-based care for patients with chronic kidney disease stages 3-5. In the UK, albuminuria testing among patients with chronic kidney disease stages 3–5 remains uncommon in primary care, with fewer than 25\% undergoing uACR testing within one year overall, but increasing to about 37\% among those formally registered with chronic kidney disease, indicating substantially higher testing conditional on chronic kidney disease recognition \cite{fraser2015timeliness}. More recent national audits report annual testing in around 30\% of patients with chronic kidney disease stages 3–5 \cite{nitsch2017national}. Similar patterns have been reported in the US, where albuminuria testing remains uncommon among adults at risk for chronic kidney disease, with ACR recorded in around 17\% of these patients, while being associated with a higher prevalence of chronic kidney disease treatment \cite{chu2023estimated}. More generally, a recent systematic review and meta-analysis of 59 studies across 24 countries, including over 3 million patients with chronic kidney disease, showed that while 81.3\% of patients received eGFR monitoring, only 47.4\% underwent albuminuria testing \cite{ketema2025quality}.\\

The example of the KFRE suggests alternative scenarios, illustrated in panels B and C of Figure 1. Panel B refers to a situation in which a set of predictors display different patterns of missingness: a measured predictor ($P_1$) serves as a proxy for unobserved underlying disease severity. As predictor $P_1$ deteriorates, further clinical investigation is undertaken, leading to measurement of predictor $P_2$, thereby restricting computation of the prognostic score to patients with recorded $P_2$. As in panel A, this situation leads to classical selection bias. Adding a direct causal relation between underlying disease severity and the measurement of $P_2$—as shown in panel C—makes the availability of $P_2$ a collider between $P_1$ and underlying disease severity.

When the situation reduces to classical selection bias, prognostic model development may still yield coefficients representative of the underlying higher-risk population. By contrast, conditioning on a collider distorts associations between all baseline predictors—not only $P_1$ and $P_2$—and the outcome \cite{greenland2003quantifying}.\\

In the context of the KFRE, declining eGFR prompts uACR testing, while perceived overall kidney failure risk—reflected by symptoms and comorbidities such as diabetes—independently influences the same decision. This double dependence of predictor availability on both eGFR and the perceived risk of the outcome characterises collider restriction bias \cite{cole2010illustrating}.

Beyond consequences regarding the applicability of prognostic models \cite{efthimiou2024developing}, patterned availability of predictors raises broader methodological issues in model development and validation, and supports consideration of simplified prognostic models \cite{WyattAltman1995,Steyerberg2004simplifying,sachs2020aim} in the setting of prospective cohorts \cite{Moons2009a}.  Taken together, these considerations highlight predictor availability as a central, yet often implicit, assumption underpinning the development, validation, and use of prognostic models.




\break
\break
\restoregeometry


\bibliographystyle{unsrt}

\end{document}